\documentclass[onecolumn, a4paper, notitlepage]{revtex4-1}
\usepackage[centering,hmargin=3.5cm,vmargin=2.5cm]{geometry}

\pdfoutput=1
\usepackage{amsmath}    
\usepackage{graphicx}   
\usepackage{verbatim}   
\usepackage{color}      
\usepackage{subfigure}  
\usepackage[hidelinks]{hyperref}   
\usepackage[percent]{overpic}
\usepackage{bm, upgreek}

\newcommand{\mum}{$\upmu$m }
\newcommand{\mumN}{$\upmu$m}

\begin{document}
	
	\title{Invited Article: Swept-Wavelength Mid-Infrared Fiber Laser for Real-Time Ammonia Gas Sensing}
	\author{R. I. Woodward$^*$, M. R. Majewski, D.~D.~Hudson and S. D. Jackson}
	\affiliation{MQ Photonics, School of Engineering, Macquarie University, New South Wales 2109, Australia}
	
\date{\today}

\begin{abstract}

The mid-infrared (mid-IR) spectral region holds great promise for new laser-based sensing technologies, based on measuring strong mid-IR molecular absorption features.
Practical applications have been limited to date, however, by current low-brightness broadband mid-IR light sources and slow acquisition-time detection systems.
Here, we report a new approach by developing a swept-wavelength mid-infrared fiber laser, exploiting the broad emission of dysprosium and using an acousto-optic tunable filter to achieve electronically controlled swept-wavelength operation from 2.89 to 3.25~\mum (3070--3460~cm$^{-1}$).
Ammonia (NH$_3$) absorption spectroscopy is demonstrated using this swept source with a simple room-temperature single-pixel detector, with 0.3~nm resolution and 40~ms acquisition time.
This creates new opportunities for real-time high-sensitivity remote sensing using simple, compact mid-IR fiber-based technologies. 

	\vspace{0.4cm}
	
	$^*$ robert.woodward@mq.edu.au
\end{abstract}

\maketitle

\section{Introduction}

The mid-infrared (mid-IR) spectral region, spanning wavelengths from 2.5 to 25~\mumN, is a frontier for photonic technology, with great potential for a range of novel applications. 
In particular, due to the existence of strong fundamental rotational-vibrational molecular absorption lines in the mid-IR, there are promising opportunities to develop new compact sensing systems---e.g. for disease diagnosis by breath analysis and stand-off detection of remote chemical threats~\cite{Hibbard2011, Bauer2008, Cossel2017}.

Critical to achieving this vision are broadband mid-IR light source and detection technologies: areas where mid-IR devices typically lag behind the performance of their near-IR counterparts.
While numerous sensing systems have been demonstrated at near-IR wavelengths (0.7--2.5~\mumN) based on vibrational overtones and combination tones, these absorption features are orders of magnitude weaker than the fundamental absorptions in the mid-IR, ultimately limiting detection sensitivity~\cite{Cossel2017, Cousin2006, Hult2007a, Millot2015}.

Conventionally, mid-IR sensing employs a thermal light source (e.g. a heated silicon carbide rod `globar') and a spectrally resolved high-sensitivity detector (e.g. a grating-based spectrometer or Fourier transform interferometer, FTIR).
This approach has limited prospect for remote sensing applications or deployment in resource-limited environments, however, due to the low coherence and low brightness of the source. 
Additionally, the need for spectrally resolved detection with mechanically moving optics limits the system robustness, increases costs, and slows the acquisition time, preventing real-time sensing.
Therefore, the development of high-brightness broadband sources and simpler detection systems for the mid-IR is an active research field.

Recent advances in source development have included remarkable demonstration of multiple-octave-spanning supercontinuum generation (e.g. using nonlinear fibers~\cite{Hudson2017, Martinez2018b, Yao2018, Salem2015} or crystals~\cite{Silva2012a, Zhou2016a} pumped with ultrafast lasers).
Despite exceeding the brightness of typical globars, the power spectral densities of reported mid-IR supercontinua are still relatively low (typically $<$-10~dBm/nm).
Higher spectral brightness can be achieved by limiting the spectral broadening to a particular region of interest (e.g. through dispersion tailoring of the nonlinear fiber), but supercontinuum detection still requires complex and costly spectrum-resolving detectors, raising a barrier to their widespread deployment.

One nascent solution to simplify detection is dual-comb spectroscopy (DCS)~\cite{Coddington2016}.
Here, two frequency comb sources with slightly different repetition rates are used, such that their interference on a simple single-pixel photodetector generates an RF heterodyne signal.
Gas absorption features which affect the optical comb lines can then be simply measured from the RF comb lines using low-cost electronics.
Impressive mid-IR DCS gas sensing performance with microsecond-scale acquisition times has recently been achieved, based on nonlinear frequency conversion of near-IR combs to the mid-IR~\cite{Coddington2016, Zhang2013n, Ycas2018}.
However, this simplification of the detection system comes at the cost of significantly more complex light sources.

Here, we propose a novel approach to mid-IR spectroscopy based on a mid-infrared wavelength-swept fiber laser, simplifying both the light source and detector requirements.
By identifying a gain material with a broad mid-IR gain bandwidth and using a fast electronically tunable intracavity spectral filter, it is possible to continuously tune the laser wavelength over a large spectral region of interest at high speed.
A low-cost single-pixel photodetector is used to continuously measure the light after a gas sample, which enables direct measurement of the spectrum by synchronizing the detector readout with the wavelength sweep rate of the source (i.e. effectively mapping the spectrum to the time domain, which can be measured on a low-bandwidth oscilloscope, eliminating complex spectrometers / FTIR setups).

Swept-wavelength lasers are mature spectral characterization tools in the near-IR, based on semiconductor or fiber amplifiers~\cite{Wysocki1990, Yun1997, Geng2011, Tokurakawa2014, YunBook2008}, although extending the concept to the mid-IR is complicated by limited availability of appropriate gain media.
While nonlinear parametric down-conversion of near-IR tunable lasers~\cite{Petrov2012}, swept sources~\cite{Silva2012} or frequency combs~\cite{Schliesser2012} is possible, as well as exploring bulk gain materials such as Cr:ZnS~\cite{Martyshkin2017}, for practical applications, more compact sources are desirable.
It should be noted that quantum cascade lasers (QCLs) and interband cascade lasers (ICLs) are a promising option with a small footprint~\cite{Phillips2014a}, although their performance at shorter mid-IR wavelengths (e.g. $<$4~\mum) is currently more modest, with reduced power scalability compared to bulk/fiber sources~\cite{Razeghi2013}.
Therefore, we target this challenging spectral region in this work using dysprosium-doped ZBLAN fiber, which is currently emerging as an ideal platform for highly efficient, compact mid-IR source development, extending the well-established benefits of fiber laser technology from the near-IR into the mid-IR.

\section{Dysprosium Spectroscopy \& Modeling}

\begin{figure}[tpb]
	\includegraphics{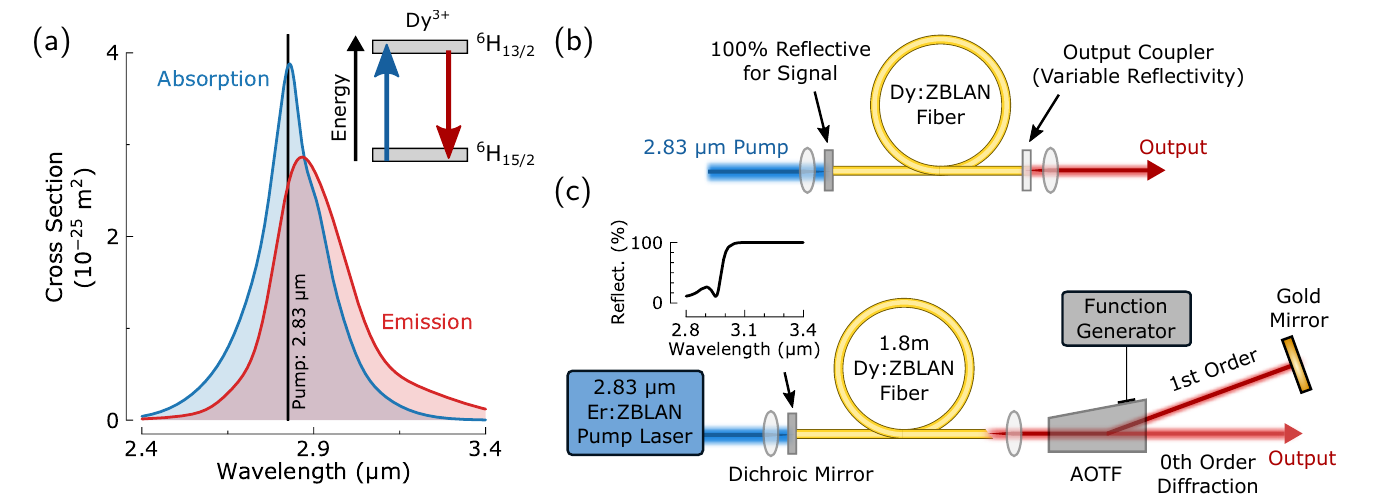}
	\caption{(a) Dysprosium-doped ZBLAN cross sections (inset: transitions shown on simplified energy level diagram). (b) Simulated linear cavity. (c) Experimental Dy:ZBLAN fiber laser cavity, including acousto-optic tunable filter (AOTF) (inset: reflectivity of input dichroic mirror).}
	\label{fig:cavity}
\end{figure}

Dysprosium (Dy) exhibits a broad emission cross section spanning from 2.5 to 3.5~\mum [Fig.~\ref{fig:cavity}(a)] (a range that includes absorption features of multiple molecular moieties e.g. OH, NH \& CH)~\cite{Jackson2003a, Gomes2010} and is currently receiving renewed research interest as a mid-IR gain material~\cite{Majewski2016, Majewski2018, Woodward2018_fsf, Woodward2018_watt, Quimby2017, Sojka2018}.
Notably, this has recently enabled the demonstration of continuously tunable CW operation over 600~nm using a rotating diffraction grating~\cite{Majewski2018} and even picosecond mode-locked operation with over 300~nm tunability~\cite{Woodward2018_fsf}.
As the laser transition is quasi-three-level and ground-state terminated, the tuning range depends critically on the doped fiber length and cavity Q-factor.
Therefore, we performed a numerical analysis to explore optimum parameters for a broadband sensing system, targeting ammonia gas (NH$_3$, with distinctive features around 3~\mumN) detection as a proof of principle.
A linear cavity design with 2000 mol.\ ppm Dy:ZBLAN fiber (12.5~\mum core diameter, 0.16 NA) is chosen [Fig.~\ref{fig:cavity}(b)], in addition to a 2.83~\mum in-band pump source, which has recently been shown to enable high slope efficiencies exceeding 70\%~\cite{Woodward2018_watt}.

Rate equation modeling is performed using measured cross sections and spectroscopic parameters~\cite{Gomes2010}. 
As our laser is in-band pumped and there are no known excited state absorption (ESA) or energy transfer upconversion (ETU) transitions for these wavelengths~\cite{Gomes2010}, the atomic system can be modeled as a simple two-level system with ground state ($^6H_{15/2}$) population $N_0$ and excited state ($^6H_{13/2}$) population $N_1$.

The power evolution $P^+(\lambda, z)$ and $P^-(\lambda, z)$ along the doped fiber (with longitudinal co-ordinate $z$) in a given direction ($+$ forwards, $-$ backwards) for each spectral channel of wavelength $\lambda$ is governed by:
\begin{equation}
\frac{\mathrm{d}P^\pm(\lambda, z)}{\mathrm{d}z} = \pm \left(
	P^\pm(\lambda, z) \left[
		g(\lambda, z) - l
	\right]
 \right)
\end{equation}
where the gain is:
\begin{equation}
g(\lambda, z) = \Gamma(\lambda) \left[
\sigma_{10}(\lambda) N_1(z) - \sigma_{01}(\lambda) N_0(z)
\right]
\end{equation}
and where $\Gamma$($\lambda$) is the overlap factor of the guided mode with the doped core, $\sigma_{10}(\lambda)$ is the emission cross section, $\sigma_{01}(\lambda)$ is the absorption cross section and $l$ is the background loss (measured as 0.3~dB/m at 3.39~\mum and assumed constant over the lasing range).

The atomic level populations $N_0$ and $N_1$ are governed by rate equations at each $z$ position:
\begin{multline}
\frac{\mathrm{d}N_1(t)}{\mathrm{d}t} = -N_1(t) \left( \frac{1}{\tau} + 
\sum_{\lambda} \frac{\sigma_{10}(\lambda) \Gamma(\lambda) [P^+(\lambda) + P^-(\lambda)]}{A_\text{core} \times h c / \lambda}
\right) 
+ \\
N_0(t) \left(
\sum_{\lambda} \frac{\sigma_{01}(\lambda) \Gamma(\lambda) [P^+(\lambda) + P^-(\lambda)]}{A_\text{core} \times h c / \lambda}
\right)
\end{multline}
where $\tau=650$~$\upmu$s is the upper state lifetime~\cite{Gomes2010}, $A_\text{core}$ is the geometric core area, $h$ is Planck's constant, $c$ is the speed of light, and $N_0$ and $N_1$ sum to equal the total doping concentration.
These equations form a boundary value problem, which is solved numerically for the steady state $\mathrm{d}N_1/\mathrm{d}t=0$ using a fourth-order collocation algorithm to identify a self-consistent evolution of power values along the fiber, subject to chosen boundary conditions (i.e. cavity mirror reflectivity values).
For our simulations, the front mirror was assumed 100\% reflective and the back mirror reflectivity was varied from 5\% to 90\%.

\begin{figure}[htpb]
	\includegraphics{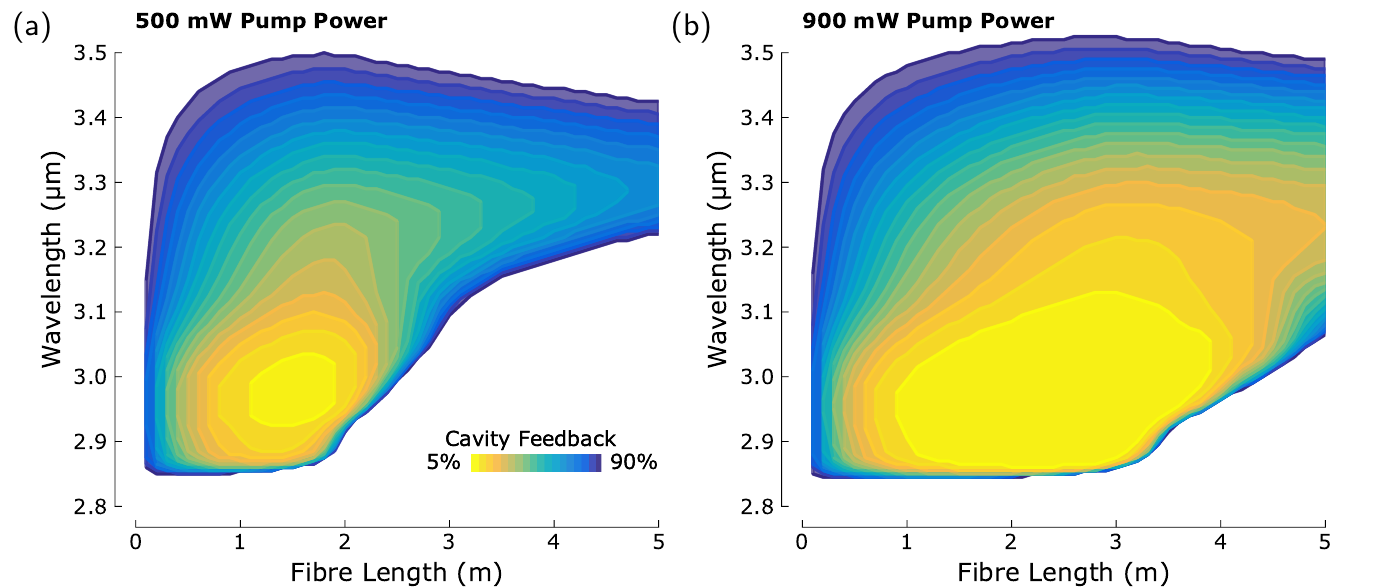}
	\caption{Simulated laser tuning range as a function of fiber length and cavity reflectivity for (a) 500~mW and (b) 900~mW pump power. Shaded regions indicate where lasing is achieved for the given operating parameters.}
	\label{fig:sim}
\end{figure}

Fig.~\ref{fig:sim} shows the tuning range computed from this model as a function of fiber length and cavity feedback ratio, where the shaded regions indicate when the applied pump power exceeded the computed laser threshold.
These simulations thus offer insight into laser design criteria to maximize tunability for in-band-pumped Dy lasers.

Firstly, it can be seen that high feedback ratios (i.e. lower cavity losses) permit wider tunability: for example, with a 1~m fiber length, the tuning range increases from 2.91--3.03~\mum to 2.85--3.46~\mum as the feedback ratio is increased from 10\% to 90\%.
This is explained by noting that towards the edges of the dysprosium emission spectral profile, the lower cross section values indicate that lower relative gain can be achieved here, therefore requiring lower cavity loss in order to achieve lasing for a given pump power.
The short-wavelength edge here is also ultimately limited by the in-band pump wavelength of 2.83~\mumN, as shown by the sharp cut-off in Fig.~\ref{fig:sim}, although even shorter wavelengths would be achievable using a non-in-band pump scheme~\cite{Majewski2018}.

With 500~mW pump [Fig.~\ref{fig:sim}(a)], increasing the amplifier length up to $\sim$1.5~m results in increased tunability due to more complete absorption of the pump power, thus maximizing the gain. For even longer lengths, however, the tuning range decreases with particular rapid decline at the short wavelength edge. 
This is due to re-absorption as the overlap of absorption and emission cross sections is particularly strong in dysprosium: the rear end of the doped fiber is not saturated as the pump has already been mostly absorbed by this point, thus generated signal light will be absorbed here and not emitted.
While re-absorption is negligible for wavelengths beyond $\sim$3.3~\mumN, the 0.3~dB/m linear loss results in gradually reduced net gain even at long wavelengths for increasing length.

At higher pump power [Fig.~\ref{fig:sim}(b)], the tuning is significantly enhanced, due to higher gain at all wavelengths.
The trend of decreasing tunability with an overly long fiber is still apparent, although the short-wavelength decline starts at a longer length of 3~m, since the increased pump power is able to saturate greater distances of fiber.

\section{Experimental Results \& Discussion}

The laser was constructed experimentally, as shown schematically in Fig.~\ref{fig:cavity}(c), with 1.8~m single-mode Dy:ZBLAN fiber (with the same specification as used for simulations, fabricated by Le Verre Fluor\'{e}).
This length was chosen to offer maximum tunability for the available pump power, at a wide range of feedback ratios.
The system also includes a butt-coupled input dichroic mirror and an external cavity at the distal end (with an 11~mm focal length black diamond aspheric lens after an angled cleave) including an acousto-optic tunable filter (AOTF, Gooch \& Housego).
The AOTF is a traveling-wave modulator employing an RF acoustic transducer coupled to a tellurium dioxide (TeO$_2$) crystal in a quasi-collinear slow-shear mode arrangement~\cite{Ward2018}, where the filter bandwidth is $\sim$5~nm and the center wavelength is set by the frequency of an applied RF sinusoid from a function generator.
Finally, the laser resonator is closed by feeding back the AOTF-diffracted light using a gold mirror, and the undiffracted light is taken as the output, where the output coupling ratio can be varied by adjusting the applied RF power, since this changes the diffraction efficiency.
A major benefit of this design is the ability to electronically tune the wavelength by varying the applied RF frequency, with no moving parts.

Initially, the AOTF is driven with a fixed 18.7~MHz RF sinusoid, centering the filter at 3.0~\mumN, and the applied RF power is set empirically to yield a maximum diffraction efficiency of $\sim$75\% (i.e. to achieve maximum cavity feedback), thus offering a 25\% output coupling ratio.
Unfortunately, the external cavity also introduces a number of losses, including 4\% parasitic reflections at the fiber tip, 15\% loss in the AOTF (per pass) and 20\% predicted coupling loss back into the fiber.
This sets an upper bound of $\sim$30\% for the cavity reflectivity.

In practice, as the pump power is increased, lasing is observed at a launched power threshold of 420~mW.
The output spectrum [Fig.~\ref{fig:basic_tuning}(a)] shows a single peak with 0.3~nm 3-dB bandwidth.
Due to the birefringent operation of the AOTF, the output is linearly polarized with 24~dB polarization extinction ratio (PER).
A slope efficiency of 33\% is recorded, which is lower than previously reported for in-band pumped dysprosium experiments~\cite{Woodward2018_watt} due to the greater cavity loss here, introduced by the AOTF.
The addition of this component, however, offers significant electronically controlled tunability.

\subsection{Electronically Tunable Dy:ZBLAN Fiber Laser}
\begin{figure}[tbp]
	\includegraphics{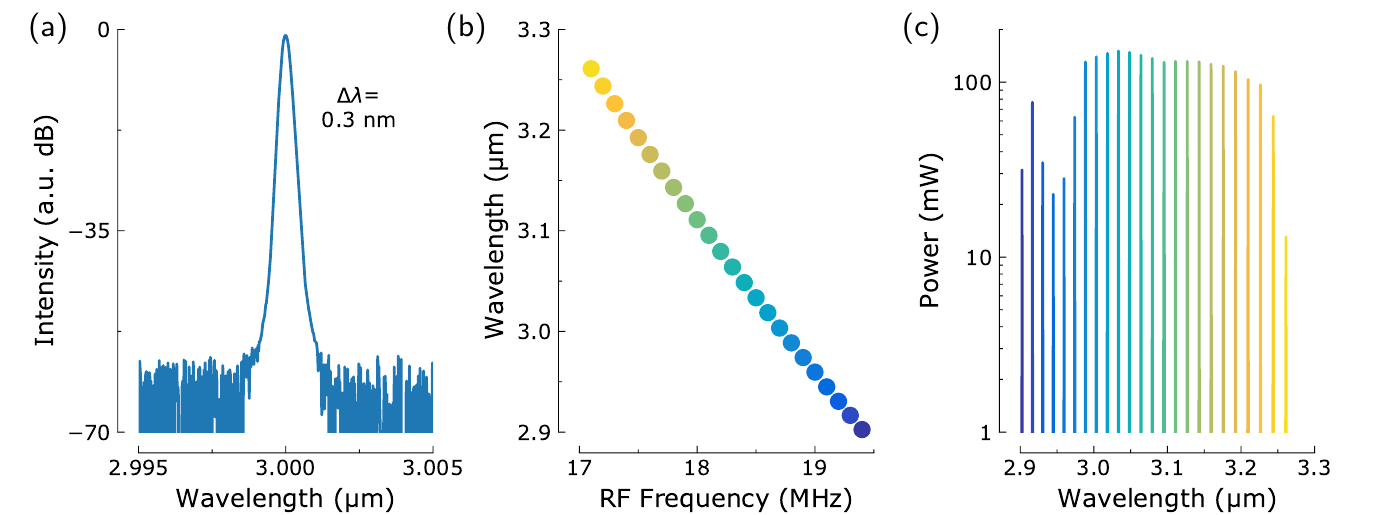}
	\caption{Tunable Dy:ZBLAN laser: (a) optical spectrum with 18.7~MHz AOTF drive frequency; (b) laser wavelength as a function of AOTF drive frequency; (c) output spectra at various RF frequencies (indicated by 
		color), showing narrowband lasing peaks, where the peak height has been normalized to the measured output power.}
	\label{fig:basic_tuning}
\end{figure}

By varying the AOTF drive frequency, the central wavelength of the filter is shifted in-situ.
With 900~mW pump power, this enables laser tunability from 2.90 to 3.26~\mumN.
The one-to-one monotonic mapping between RF frequency and laser wavelength enables simple determination of the required AOTF drive signal to achieve any desired wavelength in this range [Fig.~\ref{fig:basic_tuning}(b)-(c)].
Such random access functionality was also tested experimentally: target wavelengths were chosen from the range and the required RF frequency was obtained by interpolation of data in Fig.~\ref{fig:basic_tuning}(b).
Measurement of the resulting laser wavelength showed excellent agreement with the target wavelength, confirming repeatable spectral accuracy to within the 0.1~nm resolution of our optical spectrum analyzer.

Output powers exceeding 100~mW are achieved over the 2.99--3.22~\mum range with only small power variation with wavelength [Fig.~\ref{fig:basic_tuning}(c)]. 
At longer wavelengths, the output power falls sharply until lasing ceases beyond 3.26~\mumN, due to the decreasing emission cross section yielding lower gain.
The variation at the short wavelength edge shows a decrease in output power from 3.00 to 2.95~\mumN, but then an increase up to $\sim$2.92~\mumN, before reducing again to a loss of lasing below 2.90~\mumN.
This trend can be explained by the reflectivity profile of the input dichroic mirror [Fig.~\ref{fig:cavity}(c) inset]: while this mirror was specifically selected to separate pump and signal wavelengths, the edge is not sharp and a small ripple is observed from 2.9 to 3.0~\mumN.

For a total cavity feedback ratio of 30\% with this length of fiber, our simulations suggested a tuning range of 2.85 to 3.28~\mum could be achieved [Fig.~\ref{fig:sim}(b)].
This is in reasonably good agreement with our experimental results, particularly at the long wavelength edge, and the reduced tunability at shorter wavelengths can be related to our non-ideal input dichroic mirror that exhibited reducing reflectivity below 3~\mumN.
Therefore, in future, even greater tuning could be expected using an optimized mirror, in addition to obtaining improved anti-reflection coatings on the AOTF to lower the cavity loss.
While chromatic lens aberration does not appear to be a limiting factor here, an additional future improvement would be to replace the lens with an off-axis parabolic mirror.

\subsection{Swept-Wavelength Operation}
\begin{figure}[t]
	\includegraphics{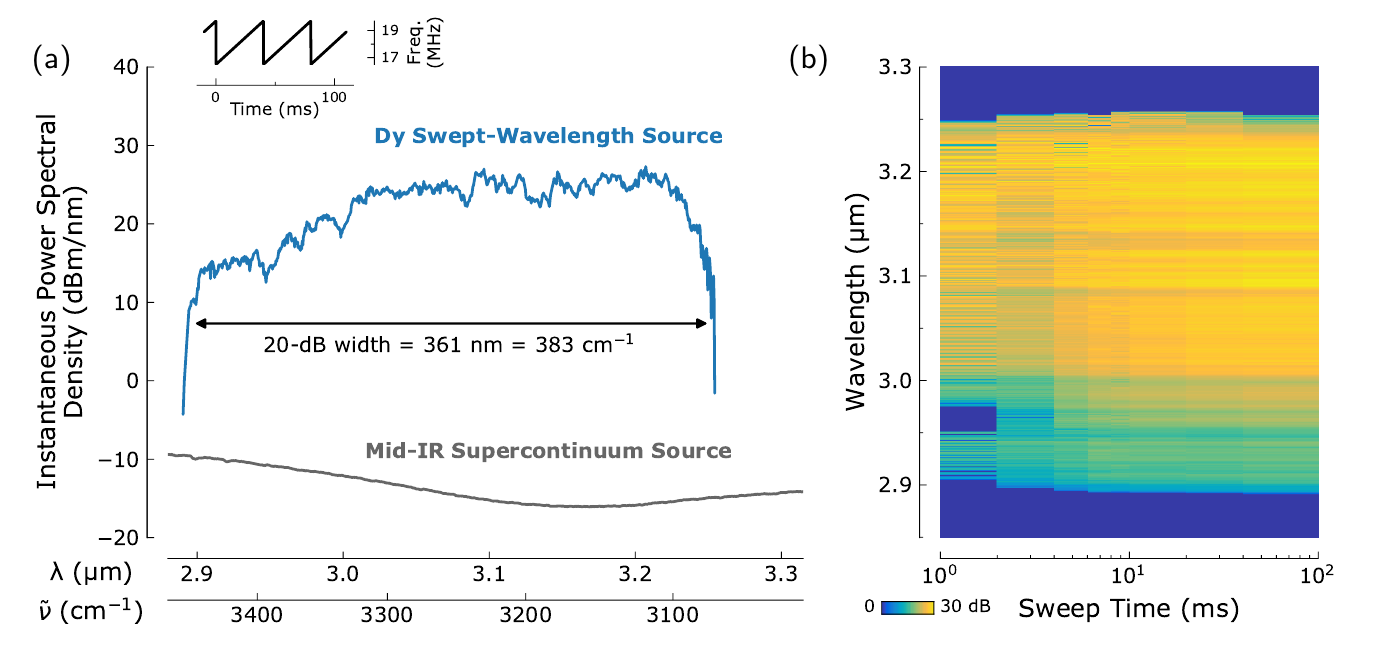}
	\caption{(a) Swept-wavelength Dy:ZBLAN peak-hold laser spectrum, shown on wavelength ($\lambda$) and wavenumber ($\widetilde{\nu}$) axes (inset: frequency chirp, with 40 ms sweep time, applied to the function generator driving the AOTF). A typical mid-IR supercontinuum spectrum (from Thorlabs ~\cite{Salem2015}) is shown for comparison. (b) Swept-wavelength peak-hold spectra as a function of sweep duration.}
	\label{fig:swept}
\end{figure}

It is also possible to operate the cavity as a swept-wavelength laser by applying a fast periodic frequency sweep to the AOTF drive signal, such that the filter central wavelength is continually varied.
Fig.~\ref{fig:swept}(a) shows the measured output spectrum for 40~ms sweep time over the range 16.5--19.7~MHz (i.e.\ 25~Hz sweep rate, shown schematically in the figure inset), using an optical spectrum analyzer in `peak-hold' mode, which measures over many sweeps.
At any instant, narrowband lasing is achieved with high power spectral density (10s--100~mW power generated within 0.3~nm bandwidth), but when the maximum spectral intensity value is recorded over a time period of a few seconds with peak-hold operation [Fig.~\ref{fig:swept}(a)], a broadband spectrum is observed from 2.89--3.25~\mumN, corresponding to 3070 to 3460 cm$^{-1}$.
We also confirmed the long-term stability of the source by noting negligible change in output spectrum after continuous operation for a number of hours.

Our instantaneous spectral brightness values of up to $\sim$27~dBm/nm (500~mW/nm) significantly exceed those of available mid-IR supercontinuum sources (typically exhibiting $<-10$ dBm/nm) in this region by many orders of magnitude [Fig.~\ref{fig:swept}(a)].
While mid-IR supercontinua naturally offer broader coverage [e.g. 1.3 to 4.5~\mum for the comparison supercontinuum source shown in Fig.~\ref{fig:swept}(a)~\cite{Salem2015}], for sensing applications where the swept-wavelength laser can cover absorption features of the target species, this approach may offer improved sensitivity and more flexible remote-detection geometries, due to the higher brightness.
Additionally, swept-wavelength sources offer simplified detection, since a simple time-resolved single pixel detector can be used due to the one-to-one mapping between time and wavelength.
To our knowledge, this is also the first reported swept-wavelength mid-IR fiber laser, and the most broadband swept fiber laser to date. 

An important question concerns the maximum sweep speed, since this determines the minimum acquisition time when using a swept-wavelength source for sensing.
The sweep rate is ultimately limited by the time taken to initiate lasing from amplified spontaneous emission (ASE), which depends on the gain material lifetime, pump power and the cavity design. 
For example, in the near-IR region, semiconductor-based swept sources have been demonstrated with MHz sweep rates, permitted by the sub-nanosecond gain dynamics of semiconductor materials~\cite{YunBook2008}.
Solid-state gain materials such as Ti:sapphire and rare-earth-doped fiber have longer lifetimes on $\upmu$s--ms time scales, however, typically leading to reduced tuning ranges and strong intensity noise from relaxation oscillations when swept faster than 100s Hz~\cite{Kodach2008, YunBook2008}.

We experimentally explored this effect by varying the duration of a single sweep from 1~ms to 100~ms (i.e. 1~kHz to 100~Hz sweep rate), generating the output spectra in Fig.~\ref{fig:swept}(b).
Above $\sim$8~ms sweep durations, there are negligible changes in the swept-wavelength spectra.
However, the tuning range does fall as the sweep time is reduced below 8~ms and the output exhibits increased intensity noise when observed on an oscilloscope, due to insufficient time for steady-state lasing to build up for the lower gain regions~\cite{Kodach2008}.
At 1~ms sweep time, the system does not lase for 2.95--2.97~\mumN, where this region relates to the low-reflectivity ripple of the input dichroic mirror, as discussed earlier.
Qualitatively, we note that reducing the sweep time of the swept-wavelength laser below 8~ms has a similar effect on tuning range as reducing the pump power.

Empirically, we discovered techniques to maximize the tuning range such as limiting the RF frequency coverage of the sweep to a range where lasing was always achieved.
When the sweep range exceeds the laser gain bandwidth, ASE operation is observed at the spectral extents, which corresponds to very different inversion dynamics compared to a lasing state: therefore, when the filter is swept back into the gain bandwidth, the required change in inversion to achieve lasing may be large. 
By contrast, if the sweep is set to just avoid ASE, the required change in inversion for lasing at one wavelength to another is likely smaller, minimizing the time required for this process and thus enabling slightly greater tuning ranges for faster sweeps.
However, for practical real-time sensing, we note that 10s~ms acquisition times are sufficient, and thus do not explore this effect further.

\subsection{Ammonia Gas Sensing}
Finally, we report a proof-of-principle application using the mid-IR swept-wavelength source for sensing of ammonia gas.
Ammonia is an important biomarker of kidney health in exhaled breath~\cite{Hibbard2011}, but is also toxic when concentrated, leading to demand for remote detection systems for environmental monitoring.

\begin{figure}[tbp]
	\includegraphics{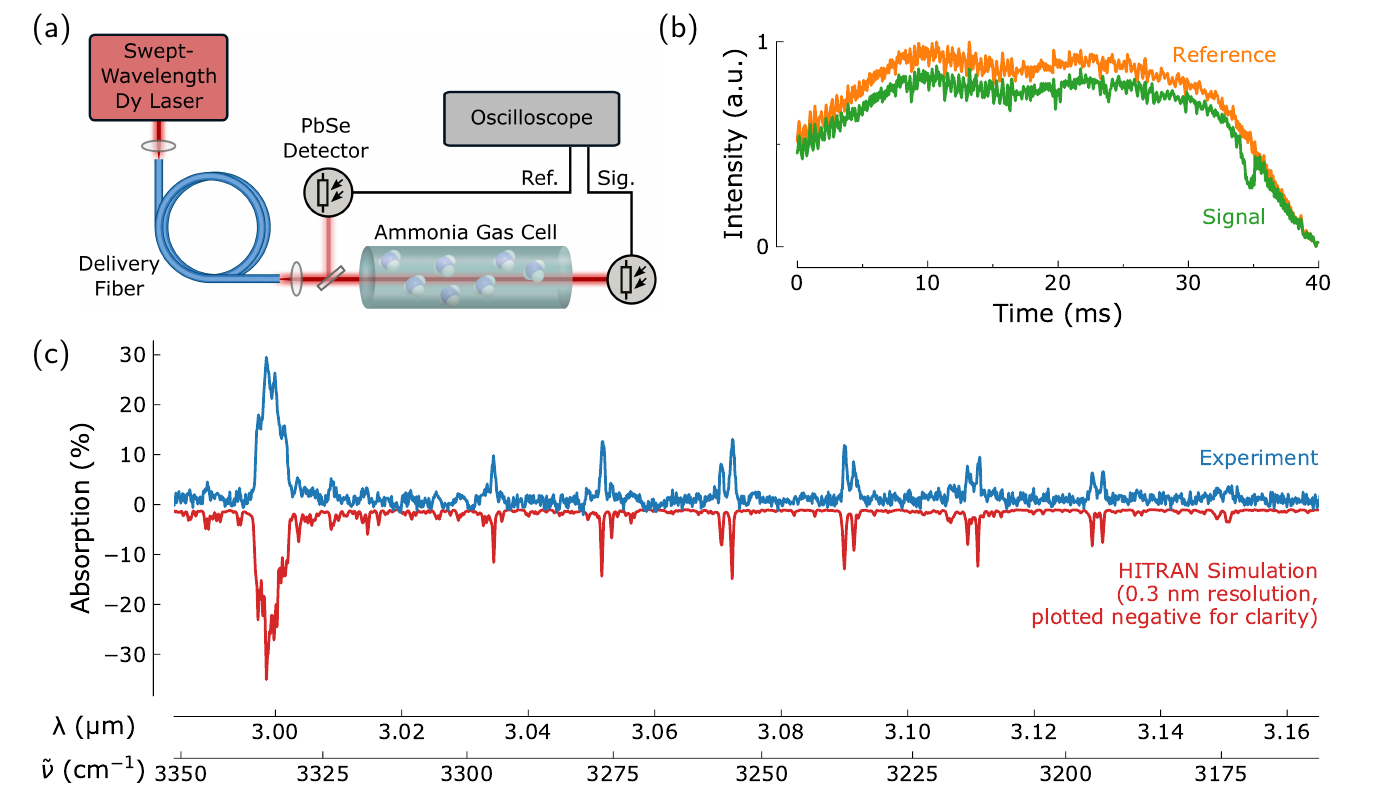}
	\caption{Ammonia gas sensing using the swept-wavelength Dy:ZBLAN fiber laser: (a) experimental setup; (b) photodetector traces recorded on oscilloscope; (c) processed measurement showing NH$_3$ absorption spectrum (compared to simulated HITRAN data).}
	\label{fig:gas}
\end{figure}

Experimentally, the swept-wavelength laser output is coupled into a single-mode passive ZBLAN fiber for beam delivery to the target [Fig.~\ref{fig:gas}(a)] (after a dichroic filter to block any residual pump light).
The high brightness and beam quality of the fiber laser platform is advantageous for convenient fiber transport of light.
The fiber output is collimated and propagated through a gas cell (with tilted, wedged MgF$_2$ windows to avoid etalon effects) containing 2\% NH$_3$ (in a nitrogen environment) at 1 atm pressure, and the instantaneous intensity of transmitted light is recorded using a low-cost room-temperature PbSe photoconductive detector. 
To eliminate common-mode noise, a fraction of the laser output is directed to a second identical photodetector for normalization.

A 40~ms sweep duration (25~Hz update time) is chosen, yielding broad tunability and fast enough acquisition for `real-time' monitoring.
The photodetector signals are digitized on an oscilloscope [Fig.~\ref{fig:gas}(b)], normalized, and the time axis is converted to wavelength based on the known sweep function applied on the function generator (mapping RF frequency to time) and the measured laser wavelength with respect to RF frequency [Fig.~\ref{fig:basic_tuning}(b)]. 
The resulting absorption measurement [Fig.~\ref{fig:basic_tuning}(c)] exhibits clear features that correspond to resonances in NH$_3$.
Simulated absorption data from the HITRAN database~\cite{Kochanov2016} is also plotted.

Excellent agreement is observed in terms of the absorption feature positions, highlighting that the system could be used to identify the presence of ammonia in an otherwise unknown environment.
The setup could be also conveniently adapted, for example, for wider area ammonia detection by propagating the collimated beam over the region of interest and retro-reflecting the light back to a detector placed near the source~\cite{Deutsch2014}.

It should also be noted that under certain conditions, swept-wavelength lasers can emit a pulsed output state with repetition rate corresponding to the cavity round-trip time---i.e. at MHz frequencies~\cite{Yun1997, Eigenwillig2013a}.
In addition, even for non-swept lasers including an AOTF driven by a constant RF sine frequency, a `frequency shifted feedback' mechanism can result in mode-locked-like picosecond pulse generation~\cite{Woodward2018_fsf}.
While such temporal modulations could add confusion to the determination of spectral features based on recording the output power as a function of time, this does not affect our setup due to the use of a reference measurement for rejecting common-mode noise, and additionally, the 10~kHz bandwidth of our PbSe detectors will average out any MHz-scale fluctuations over the time scale of our absorption measurements.

The wavelength resolution is an important metric for practical sensing, which here is limited by two factors.
Firstly, the 0.3~nm ($\sim$0.3 cm$^{-1}$) spectral bandwidth of the source at any instant.
Secondly, with a similar order of magnitude, is the limited detector bandwidth $B$: the nominal 10~kHz bandwidth corresponds to $\tau=0.35/B=35~\upmu$s rise time, meaning that spectral features swept faster than this time period will be averaged out.
For the ammonia sensing demonstration (Fig.~\ref{fig:gas}), our laser swept across a 180~nm spectral span in 40 ms, suggesting a resolution limit of 0.16 nm per 35~$\upmu$s.

Precisely determining and enhancing the minimum system sensitivity, in addition to utilizing measured information in the strength of the absorption features compared to theory to determine gas pressure/temperature, are promising topics for future work, which will be achieved by upgrading the fixed gas cell to a variable flow setup.
With flowing gas, it should also be possible to characterize changes in ammonia content in real time based on the 25~Hz sweep rate.
We highlight, however, that the current system has successfully demonstrated the principle of mid-IR swept-wavelength fiber laser gas sensing for the first time, opening the door to additional investigations using this compact, flexible sensing platform.

In terms of infrared spectroscopy, we are specifically measuring absorption features corresponding to excitation of stretching vibrations of the N-H bond~\cite{Thompson2018}.
This bond is found in many compounds (e.g.\ amides and amines) and while the precise position and strength of absorption features depend on other bonds in the overall molecule, N-H stretching generally corresponds to absorption in the 3000-3400 cm$^{-1}$ region~\cite{Thompson2018}.
As a result, the broad tuning range of our swept-wavelength laser that spans this region suggests numerous further applications for different materials.
For example, stand-off detection of ammonium nitrate (N$_2$H$_4$O$_3$), which is a major constituent of explosives, is of great importance and has been shown to be enabled by broadband light sources around 3~\mumN~\cite{Kumar2012}.
Our tuning range also approaches stretching vibrations of the C-H bond (typically 2850--3000 cm$^{-1}$) paving the way to sensing of alkanes such as methane (CH$_4$, critical to quantify for environmental monitoring).

Beyond the 3-\mum dysprosium transition we demonstrate here, this swept-wavelength fiber laser concept could also be extended to other wavelength ranges by simply changing the gain medium and pump source, e.g. to Er:ZBLAN or Ho:InF$_3$ for 3.35 to 3.95 emission ~\cite{Henderson-Sapir2016a, Maes2018a}, or even Dy:InF$_3$~\cite{Majewski2018c} for emission beyond 4~\mumN.
As TeO$_2$ has a broad transparency window from 0.4 to 4.5~\mumN, TeO$_2$-based AOTFs can operate up to $\sim$4.5~\mum \cite{Ward2018}, and PbSe detectors offer reasonable responsivity up to 5~\mumN; thus, the remainder of the setup could remain unchanged.

\section{Conclusion}
In summary, we have exploited the promising spectroscopy of dysprosium-doped fluoride fiber to demonstrate the first mid-IR swept-wavelength fiber laser, achieving over 360~nm sweep range around 3~\mum at up to kHz sweep speeds.
The source was practically applied for real-time remote sensing of ammonia gas, using low cost single-pixel detectors at room temperature, showing excellent agreement between measurements and simulated spectral features.
This represents the first practical application of mid-IR fiber laser technology for gas sensing, to our knowledge, showing that fiber lasers have significant application potential beyond the near-IR and paving the way to compact low-cost optical sensing technologies suitable for deployment in real-world environments.

\section*{Funding Information}
Australian Research Council (ARC: DP170100531).

\section*{Acknowledgments}
We thank Dr Ori Henderson-Sapir for useful discussions.
RIW acknowledges support through an MQ Research Fellowship.

\end{document}